\documentstyle[aps,pre,preprint]{revtex}

\begin{document}

\draft

\title{
Lyapunov spectral analysis of a nonequilibrium Ising-like transition }

\author{
Corey~S.\ O'Hern\cite{cso-email},
David~A.\ Egolf\cite{dae-now}, and
Henry~S.\ Greenside\cite{CNCS-address}
}
\address{
Department of Physics\\
Duke University\\
Durham, NC 27708-0305
}

\date{June 17, 1995}

\maketitle

\begin{abstract}

By simulating a nonequilibrium coupled map lattice that
undergoes an Ising-like phase transition, we show that
the Lyapunov spectrum and related dynamical quantities
such as the dimension correlation length~$\xi_\delta$
are insensitive to the onset of long-range
ferromagnetic order.  As a function of lattice coupling
constant~$g$ and for certain lattice maps, the Lyapunov
dimension density and other dynamical order parameters
go through a minimum. The occurrence of this minimum as
a function of~$g$ depends on the number of nearest
neighbors of a lattice point but not on the lattice
symmetry, on the lattice dimensionality or on the
position of the Ising-like transition.  In one-space
dimension, the spatial correlation length associated
with magnitude fluctuations and the length~$\xi_\delta$
are approximately equal, with both varying linearly
with the radius of the lattice coupling.

\end{abstract}


\section{Introduction}
\label{intro}

Laboratory experiments \cite{space-time-expts} and
numerical simulations \cite{simulation-refs} can now
systematically explore sustained homogeneous
nonequilibrium systems of quite large aspect ratios
($\Gamma \lesssim 1000$) which possibly approximate a
thermodynamic limit of infinite system size.  These
advances raise the theoretical question of identifying
order parameters for analyzing and classifying
spatiotemporal chaotic states so that a quantitative
comparison can be made between theory and experiment
\cite{CrossHohenberg94}.  The most appropriate order
parameter for a given nonequilibrium system is
presently not well understood although numerous
possibilities have been studied. Some order parameters,
such as spatial correlation lengths obtained from
exponentially decaying correlation functions, emphasize
the average spatial disorder and have been widely used
in condensed matter physics \cite{Hohenberg89}.  Others
such as the metric entropy and the Lyapunov fractal
dimension \cite{Ott93} are familiar from nonlinear
dynamics and emphasize the average temporal disorder or
dynamical complexity arising from the geometric
structure of an attractor in phase space.

We would then like to know whether these different
kinds of order parameters are related and whether there
is a need for new order parameters.  As an example,
does knowledge of an easily measured correlation length
give information about the fractal dimension, which is
difficult to estimate from experimental time series
\cite{time-series-limits}?  That a relation between
spatial disorder and dynamical complexity may exist is
suggested by the prominence in many nonequilibrium
states of defects \cite{Cr93} whose dynamics often
determine the average spatial disorder
\cite{CoulletGilLega89}.  Examples of defects are
amplitude holes in one-dimensional complex
Ginzburg-Landau equation (abbreviated below as CGLE)
\cite{Nozaki84,Sakaguchi88,Shraiman92,Egolf95prl},
domain walls and droplets of opposite spin in coupled
map lattices (CMLs) with an Ising-like transition
\cite{Sakaguchi91Ising,Miller93}, vortices in the
two-dimensional CGLE \cite{CoulletGilLega89}, and
spirals and centers in the recently discovered
spiral-defect chaos state in thermal convection
\cite{SpiralDefectChaos}.  The nucleation, motion, and
annihilation of defects are important features of the
chaotic dynamics and so dynamical quantities such as
the fractal dimension may be related to their spatial
statistics \cite{Chate93,Egolf94nature,Afraimovich95}.
Complicating this simple picture is the fact that not
all fluctuations are associated with defect motion,
e.g., phase fluctuations in the CGLE
\cite{Shraiman92,Egolf94thesis}. The fractal dimension
may then be larger than that suggested by defect
statistics.

In recent work
\cite{Egolf94nature,Egolf95prl,Egolf94thesis}, Egolf
and Greenside have explored the relation between
temporal and spatial disorder for the CGLE on a large
periodic interval \cite{Shraiman92}.  Observing that
sufficiently large chaotic systems become extensive so
that a fractal dimension~$D$ grows linearly with volume
size~$V \propto L^d$ (where~$L$ is the system size
and~$d$ is the spatial dimensionality)
\cite{Cr93,Ma85,Gr89,exp-extr,Egolf94nature,vandeWater93,OHern95algdecay},
they calculated the dimension correlation
length~$\xi_\delta$ which is defined \cite{Cr93} in
terms of the intensive dimension density
\begin{equation}
  \delta = \lim_{L \to \infty} D/V , \label{delta-defn}
\end{equation}
by the equation
\begin{equation}
  \xi_\delta = \delta^{-1/d} . \label{xi-delta-defn}
\end{equation}
The length~$\xi_\delta$ can be interpreted crudely as a
characteristic size of dynamically independent
subsystems or a characteristic range of chaotic
fluctuations.
Near a
transition from phase- to defect-turbulent states
\cite{Shraiman92,Chate95santafe}, Egolf and Greenside
found that the length~$\xi_\delta$ was approximately
equal to, and had the same parametric dependence as,
the spatial correlation length given by the magnitude
fluctuations of the Ginzburg-Landau field
\cite{Egolf95prl}.  Over the same parameter range, the
correlation length~$\xi_\phi$ arising from phase
fluctuations (and also of the Ginzburg-Landau field
itself) was found to increase to quite large values,
suggesting that the chaos fluctuations measured
by~$\xi_\delta$ were short-ranged and decoupled from
the phase.

In this paper, the one-dimensional investigations of
Egolf and Greenside \cite{Egolf94nature,Egolf95prl} are
extended by examining similar questions of how spatial
disorder and dynamical complexity are related for
spatial dimensionality $d=2$ and~$d=3$.  We study a
class of dissipative coupled map lattices
\cite{Miller93} that undergoes an Ising-like
ferromagnetic-ordering transition as a lattice coupling
constant~$g$ (defined below in
Eq.~(\protect\ref{eq:Miller-Huse-map})) is increased through a
range of finite positive values, corresponding to a
transition from a high-temperature paramagnetic phase
to a low-temperature ferromagnetic phase. Instead of
the point-like space-time defects of the
one-dimensional CGLE \cite{Egolf95prl}, long-lived
defects occur in the form of domain walls and droplets
involving regions of opposite sign.  At a critical
transition point~$g=g_c$, the usual two-point
correlation length~$\xi_2$ diverges to infinity while
the magnetization (the average of the signs of all
lattice variables) bifurcates from a zero to nonzero
value. The main attraction of the Miller-Huse model
\cite{Miller93} is that a transition with a diverging
correlation length occurs from one chaotic state to
another. This permits a careful comparison of different
length scales near the transition point.

By calculating the Lyapunov spectrum and related
dynamical quantities such as the dimension correlation
length~$\xi_\delta$, we show that correlations in
chaotic fluctuations near the ordering transition have
a short range and are decoupled from the diverging
long-range order measured by the length~$\xi_2$.  As
was the case for the 1d periodic CGLE
\cite{Egolf95prl}, the length~$\xi_\delta$ and the
correlation length~$\xi^{\rm mag}_2$ arising from
magnitude fluctuations of the CML variables both turn
out to be short---here about one lattice spacing---but
a quantitative relation can not be determined since
these lengths do not vary substantially with
parameters.
In one-space dimension, a more substantial variation in
these quantities is obtained by increasing the
radius~$r$ of the coupling from nearest to
$r$th-nearest lattice neighbors.  With increasing
radius~$r$, the lengths $\xi_\delta$ and $\xi^{\rm
mag}_2$ then increase approximately linearly and with
slopes related by a factor of order one.
This suggests that the two lengths may be related and
that~$r$ is important in
determining the length scale of dynamical fluctuations.

Several dynamical quantities such as the largest
Lyapunov exponent~$\lambda_1$, the metric entropy
density~$h = \lim_{L \to \infty} H / V$,
and the Lyapunov fractal dimension density~$\delta$ attain minimum
values near---but distinctly not at---the critical
transition point~$g=g_c$.
At first glance, minima in
quantities such as~$h$ or~$\delta$ seem counterintuitive
since the onset of ferromagnetic ordering should
correspond to increased correlations, i.e., decreased
dynamical complexity and a decreased dimension density or
entropy density. Such a monotonic decrease in the metric
entropy is, in fact, observed for a non-dissipative
equilibrium CML that undergoes an Ising-like transition
\cite{Sakaguchi91Ising}.  But because the dimension
correlation length~$\xi_\delta$ is quite short in the
Miller-Huse model, we argue below that dynamical
quantities are only sensitive to nearest neighbor
dynamics. As the coupling constant~$g$ increases, the
discrete Laplacian eventually becomes antidiffusive,
magnifying rather than reducing short-wavelength
structure, and the dimension density and entropy density
start to increase.

The role of coupling lattice neighbors is demonstrated
with calculations on periodic 2d~hexagonal and 3d~cubic
lattices.  For these cases, $\xi_\delta$ is again about
one lattice spacing in size and the extrema in
dynamical quantities occur at a value~$g \approx
1/(n+1)$ determined by the number~$n$ of nearest
neighbors ($n=4$ for the square lattice, $n=6$ for the
hexagonal and cubic lattices). The positions of the
extrema do not coincide with the bifurcation of the
magnetization and do not seem to depend on the lattice
symmetry and dimensionality. The origin of these minima
remains to be explained.

These results, together with previous work on the
1d~CGLE and with some unpublished work on CMLs with
algebraic decay of spatial correlations
\cite{OHern95algdecay}, suggest that the dimension
correlation length will typically be short so that
chaotic fluctuations are decoupled from long-range
spatial order as measured by correlation functions.
This leads to the qualitative conclusion that defects,
although a striking visual feature of spatiotemporal
chaos, are not the source of complexity that leads to
large fractal dimensions and to small dimension
correlation lengths.  The physical meaning and utility
of the length~$\xi_\delta$ remains to be understood and
further studies on different kinds of systems will be
useful.

The rest of this paper is organized as follows.  In
Section~\protect\ref{methods}, we define the coupled map lattice and
discuss some details of how its Lyapunov spectrum is
calculated using a CM-5 parallel computer
\cite{Hillis93}. In Section~\protect\ref{results}, we discuss
various results of our simulations, especially the
dependence of order parameters on the lattice coupling
constant and on the symmetry and dimensionality of the
lattice. Finally, in Section~\protect\ref{conclusions}, we summarize
our results and relate them to other recent research.

\section{Methods}
\label{methods}

In this section, we define the mathematical models used
in our simulations and discuss some details about how
the Lyapunov spectrum and spatial correlation lengths
were calculated numerically. Since a CM-5 parallel
computer played an important role in our being able to
explore large space-time regions for many parameter
values, we also discuss some details of how the
algorithms were adapted for parallel computation.

\subsection{Models}
\label{models}

As easily simulated models of spatiotemporal chaos, we
consider homogeneous coupled map lattices (CMLs) in
which the same chaotic map~$\phi(y)$ is associated with
each point of a finite periodic lattice and for which
nearest neighbor maps are coupled linearly by
diffusion. CMLs have a significant advantage over
partial differential equations of being analytically
amenable \cite{Bunimovich88} and easier to simulate on
a computer.  CMLs have the drawback that their
solutions can not generally be related quantitatively
to experiment and they may not have universal critical
properties near transitions \cite{Marcq95ising}.

We study CMLs suggested by recent work of Miller and
Huse \cite{Miller93}, who analyzed the long-wavelength
properties of a two-dimensional CML that orders
ferromagnetically, in analogy to the equilibrium Ising
model \cite{Landau80}. Following these authors, we use
a lattice map~$\phi(y) = - \phi(-y)$ with odd symmetry
so that domains of opposite ``spin'' or sign arise.
The odd symmetry is a necessary but not sufficient
condition \cite{Boldrighini95} for obtaining an
Ising-like transition in which the magnetization
(defined below in Eq.~(\protect\ref{eq:magnetization-defn}))
bifurcates to a nonzero value as a coupling
constant~$g$ is varied.  Following Miller and Huse, we
choose~$\phi(y)$ to be a piecewise-linear map with
slope of constant magnitude greater than one, for which
only chaotic states of constant measure exist in the
absence of lattice coupling.

If~$y^t_i$ denotes the variable at spatial site~$i$ at
integer time~$t$ (with $t=0$, $1$, $\cdots$), then the
rule for updating each lattice variable to time~$t+1$
is given by \cite{Miller93}:
\begin{equation}
  y_i^{t+1} = \phi( y_i^t )
    + g \sum_{j(i)} \left(
          \phi( y_j^t )
        - \phi( y_i^t )
      \right)
  ,  \label{eq:the-cml}
\end{equation}
where the parameter~$g$ is the spatial coupling
constant.  The sum goes over indices~$j(i)$ that denote
nearest neighbors sites of site~$i$, e.g., the four
nearest neighbors on a square two-dimensional lattice
or the six nearest neighbors in a 2d hexagonal lattice
or 3d cubic lattice. For most of our calculations, we
used the same local map as in Ref.~\cite{Miller93}:
\begin{equation}
\phi(y) = \left\{
  \begin{array}{ccrcccr}
    -2 - 3y &  \mbox{for} &  \mbox{$-1$\ } & \le & y & \le & -1/3 ,\\
       3y   &  \mbox{for} & -1/3 & \le & y & \le &  1/3 ,\\
     2 - 3y &  \mbox{for} &  1/3 & \le & y & \le & \mbox{$1$,\ }
  \end{array}
  \right.
  \label{eq:Miller-Huse-map}
\end{equation}
with a slope of constant magnitude equal to~3.  To
understand the competition between local chaos and
diffusion (which decrease and increase spatial
correlations respectively), we also used a more weakly
chaotic map with a slope of smaller constant magnitude
equal to 1.1:
\begin{equation}
\phi(y) = \left\{
  \begin{array}{cl}
    -1.1 x - 1.0  & \mbox{for $x \le 0$,}\\
    -1.1 x + 1.0  & \mbox{for $x > 0$.}\\
  \end{array}
  \right.
  \label{eq:slope-1.1-map}
\end{equation}

Once a local map~$\phi(y)$ has been chosen, a numerical
simulation is specified by the dimensionality of the
lattice~$d$, the symmetry of the lattice (e.g., square,
hexagonal, or cubic), the size of the lattice~$L$
(number of sites along an edge), the coupling
constant~$g$, the initial condition~$y_i^0$, and the
total integration time~$T$. In this paper, we used an
integer lattice in~1d, square and hexagonal lattices in
2d and a cubic lattice in~3d. Initial conditions
consisted of assigning a random uniformly-distributed
number in the interval $[-0.1,0.1]$ to each site;
results were not dependent on the choice of initial
conditions provided the integration time was
sufficiently long.  Typical integration times
were~$T=50,000$ iterations for calculations of Lyapunov
exponents and~$T=150,000$ iterations for calculations
of correlation functions.  We made a few runs with
longer integration times of~$T=60,000$ and~$T=500,000$
to check the convergence of the Lyapunov
exponents and the correlation functions, respectively.
Different lattice sizes were used depending on which
order parameters were being studied, typically $L \le
32$ for calculating Lyapunov exponents (which are quite
costly to compute) and $L \le 1024$ for estimating
correlation lengths.  Many runs were repeated using
several lattice sizes to verify the absence of
significant finite-size effects.

For calculations of statistical averages such as the
magnetization and correlation functions, the statistics
could often be improved substantially by averaging the
results of an ensemble of~$N$ runs each of
duration~$T$, with each run differing only in the
choice of initial conditions.  Calculations indicate
that this ensemble average is ergodically equivalent to
a single integration of duration~$N T$
\cite{Egolf94thesis,Egolf95ics}.  For most of the
results reported below, we used an ensemble average
over~$N=64$ runs which could be executed simultaneously
and in parallel on the vector units of a 16-node
partition of a CM-5 computer.

\subsection{Lyapunov Exponents and Associated Dynamical Quantities}
\label{exponents}

Some dynamical order parameters can be constructed by
combining in different ways the Lyapunov
exponents~$\lambda_i$ associated with a given attractor
\cite{Ott93}.  For the CML given by Eq.~(\protect\ref{eq:the-cml})
with a total of~$N=L^d$ lattice sites, there are~$N$
real-valued Lyapunov exponents~$\lambda_i$ (labeled in
decreasing order $\lambda_1 \ge \lambda_2 \ge \cdots
\ge \lambda_N$) that characterize the long-time
average-rate-of-separation of nearby orbits in phase
space.  From the~$\lambda_i$, we can calculate a
Lyapunov dimension~$D$ given by the Kaplan-Yorke
formula \cite{Ott93}:
\begin{equation}
  D = K
    +  {1 \over |\lambda_{K+1}| } \sum_{i=1}^K \lambda_i
  , \label{eq:dim-defn}
\end{equation}
and calculate an entropy defined by the sum of the
positive exponents \cite{Ott93}:
\begin{equation}
  H = \sum_{\lambda_i  > 0} \lambda_i
  . \label{eq:entropy-defn}
\end{equation}
The number~$K$ in Eq.~(\protect\ref{eq:dim-defn}) is the largest
integer such that the sum~$\sum_{i=1}^K \lambda_i$ of
the first~$K$ exponents is nonnegative; this sum is
positive for $K=1$ if an orbit is chaotic ($\lambda_1 >
0$) and is negative for~$K=N$ if the dynamics is
dissipative and so the sum typically crosses zero at an
intermediate index~$K \approx D$ for a chaotic
dissipative system.

The exponents~$\lambda_i$ were calculated numerically
by a now-standard numerical method \cite{Parker89}, in
which $K \le N$~linearizations of the equations of
motion, Eq.~(\protect\ref{eq:the-cml}), are evolved in time. This
allows one to follow~$K$ Lyapunov vectors in a tangent
space from which local stretching information and the
Lyapunov exponents can be extracted. Together with a
particular nonlinear orbit defined by the equations of
motion, a total of $K+1$ CMLs was evolved to
calculate~$K$ Lyapunov exponents.

Repeated orthonormalizations of Lyapunov vectors at
time intervals~$T_n$ are needed to prevent
floating-point overflow from the exponentially growing
values and to prevent inaccuracies arising from the
loss of linear independence as they fold up along the
direction of the fasting growing exponent
\cite{Parker89}.  For the maps
Eqs.~(\ref{eq:Miller-Huse-map})
and~(\ref{eq:slope-1.1-map}), we found empirically that
values $T_n \lesssim 30$ gave reasonable results for
all lattices studied with the largest value of~$T_n$
depending on the parameters~$g$ and~$L$.  Smaller
renormalization times did not change the values of the
Lyapunov spectrum (although the code was more costly to
run) while larger values led to serious errors due to
linear dependence of the Lyapunov vectors.

The orthonormalizations of tangent vectors consumed
most of the computing time on a Thinking Machines CM-5
parallel computer.  The orthonormalizations require
substantial communication between different processors
since each processor evolves independently only a few
of the~$K$ linearized equations in its own local
memory; this communication decreases the efficiency of
the code.

Although the communication inherent in the orthonormalization
procedure could not be avoided,
in all other portions of the code
communication between nodes was eliminated
by iterating redundantly an identical copy of the
nonlinear CML Eq.~(\protect\ref{eq:the-cml}) with identical initial
conditions on each processor. In this way, information
about the underlying orbit (needed when iterating the
linearized CMLs) did not have to be communicated from
one processor to all others at each time step.

By monitoring the Lyapunov exponents and the Lyapunov
dimension as a function of time~$t$, we found
empirically that an integration time~$T \gtrsim 50,000$
time steps gave an acceptable relative accuracy of
better than one percent for calculating the
dimension~$D$ and entropy~$H$ for all lattices studied
($L \le 32$). Fig.~\protect\ref{fig:dim-vs-time} shows how the
dimension~$D$ converges over time for lattice
size~$L=16$ and for~$g=0.202$. The dimension curve is
noisy with fluctuations that diminish slowly over time
(the $\lambda_i$, not shown, have substantially noiser
time series). Goldhirsch et al \cite{Goldhirsch87} have
argued that the amplitude of the exponent fluctuations
should decay as~$1/T$ where~$T$ is the total
integration time and so one could fit and extrapolate
to get a better estimate \cite{exp-extr}.
Extrapolation was not needed in plots like
Fig.~\protect\ref{fig:dim-vs-time} which already give an adequate
relative accuracy of better than one percent.

By repeating plots such as Fig.~\protect\ref{fig:dim-vs-time} for
different system sizes~$L$ with all other parameters
held fixed, we found extensive scaling of the
dimension~$D$ with the volume of the system~$N=L^2$ for
a wide range of parameter values~$g$, an example of which is given
in Fig.~\protect\ref{fig:ext-chaos}.
Fig.~\protect\ref{fig:ext-chaos}(a)
shows that the dimension~$D$ increases linearly and
extensively with~$N$ beyond a system size of
about~$L=9$. The Lyapunov dimension density~$\delta$
could then be obtained from the slope of a
least-squares fitted line in the extensive region. The
intercept of a least-squares-fitted line through the
four right-most points was~$0.007$ which is quite small
(and is also approximately zero for extensively chaotic
solutions of the 1d~CGLE \cite{Egolf94thesis}). There
is then the possibility that a single dimension
calculation for a sufficiently large system may suffice
to estimate its dimension density.
By comparing the dimension per volume~$D/L^2$ with
the dimension density $\delta$,
Fig.~\protect\ref{fig:ext-chaos}(b) shows how the extensive
regime is approached rapidly and achieved for fairly
small system sizes~$L \ge 9$.

We finish this subsection with two comments about the
meaning of the Lyapunov dimension~$D$ and about why we
chose to calculate~$D$ from Lyapunov vectors rather
than from time series measurements. The reader should
recall that there is an infinity of fractal
dimensions~$D_q$ (often called the Renyi dimensions)
associated with a strange attractor with the
parameter~$q$ varying over the real numbers
\cite{Ott93}. Since each dimension~$D_q$ will be
extensive for a homogeneous extensively chaotic system,
the particular values only reflect the system size and
are not interesting themselves.  Instead, one should
define a continuum of corresponding intensive dimension
densities~$\delta_q = \lim_{L \to \infty} D_q/V$ to
provide a partial characterization of such systems.

For large fractal dimensions ($D \gtrsim 5$), present
computers and algorithms only allow the calculation of
the Lyapunov fractal dimension Eq.~(\protect\ref{eq:dim-defn}) and
its corresponding density.  The Lyapunov dimension is
conjectured to be the same as the Renyi dimension
with~$q=1$ (i.e., the information dimension~$D_1$
\cite{Ott93}) and it is not known to what extent the
corresponding density~$\delta_1$ characterizes the
unknown function of densities~$\delta_q$, e.g., whether
it is close to the mean value of the~$\delta_q$.  There
are one-dimensional maps for which the ratio of $D_1$
to~$D_2$ (the two most commonly calculated fractal
dimensions) can be arbitrarily large \cite{Cutler90},
and so the variation of the function~$\delta_q$ around
its mean value can be large.  A perhaps even more
important question is whether the different dimension
densities~$\delta_q$ each have a similar dependence on
model parameters, e.g., all increasing or decreasing
together. Until this issue is resolved, the dimension
density~$\delta_1$ and the corresponding
length~$\xi_\delta = \delta_1^{-1/d}$ need to be
interpreted with caution.

The dynamical quantities~$D$ and~$H$ are calculated in
terms of the Lyapunov spectrum~$\lambda_i$, and not in
terms of time series~$y_i^t$ at a given lattice
site~$i$, because of the impractical computational
demands of time series algorithms \cite{Abarbanel93}.
While the computational complexity of the method based
on the Lyapunov spectrum scales algebraically with
system volume~$V$ or dimension~$D$
\cite{ref:alg-scaling}, it is well known that the
computational complexity of classical time series
algorithms such as that proposed by Procaccia and
Grassberger grows {\sl exponentially} with $D$,
imposing severe restrictions on the largest dimension
that can be estimated from experimental data.  Although
it remains controversial what is the practical upper
bound for clean time series of less than a million
points (there are claims from~6 to~20
\cite{time-series-limits}), existing time series
methods can not treat extensively chaotic systems whose
fractal dimensions may be in the hundreds (see
Fig.~\protect\ref{fig:ext-chaos}).

\subsection{Magnetization and Correlation Lengths}
\label{lengths}

In addition to the dynamical quantities described in
the previous section, we quantify the evolution of the
CMLs by a ``magnetization''~$M$ and by length scales
measured from two-point and mutual information
correlation functions. We describe these briefly to
indicate the method and errors involved.

Following Miller and Huse \cite{Miller93}, the average
magnetization~$M$ of the CML is defined by a space-time
and ensemble average of the signs $\pm 1$ of the
lattice values~$y^t_i$,
\begin{equation}
  M = \langle {\rm sign}(y_i^t) \rangle
    \equiv {1 \over N T p} \sum_{i,t,p} {\rm sign}(y_i^t(p))
  , \label{eq:magnetization-defn}
\end{equation}
where the index~$p$ labels a particular CML running on
processor~$p$. An average over 64~independent CMLs
running on a 16-node partition of a CM-5 was typically
used.  For the local maps Eq.~(\protect\ref{eq:Miller-Huse-map})
and Eq.~(\protect\ref{eq:slope-1.1-map}), $M$ undergoes a pitchfork
bifurcation from a zero to finite value as the lattice
coupling~$g$ is increased from small values. The
critical value~$g_c$ at which~$M$ bifurcates to a
nonzero value is unchanged if the values of the site
variables~$y_i^t$ are used instead of their signs in
Eq.~(\protect\ref{eq:magnetization-defn}). The bifurcation of~$M$
to a nonzero value defines the onset of ferromagnetic
order at $g=g_c$ \cite{Miller93} as illustrated in
Fig.~\protect\ref{fig:M-for-square-lattice}.

To characterize the average spatial disorder, we
examined two of many possible definitions of spatial
correlation lengths, one from an exponentially decaying
two-point correlation function, another from an
exponentially decaying mutual information function
\cite{mutual-info-refs}. (Some other correlation
lengths are discussed on pages~945-947 of
Ref.~\cite{Cr93}.)  The two-point correlation function
was defined in the usual way:
\begin{equation}
  C_2(|{\bf X}_i - {\bf X}_{i'}|) =
  \left\langle
    ( y^t_i  -  \langle y \rangle )
    \,
    ( y^t_{i'} -  \langle y \rangle )
  \right\rangle
  , \label{eq:2-pnt-correlation-fn}
\end{equation}
where ${\bf X}_i$ denotes the position of lattice
point~$i$ and where the brackets $\langle \cdot
\rangle$ denote the averaging process of
Eq.~(\protect\ref{eq:magnetization-defn}). Given the periodicity of
the lattice and the availability of efficient parallel
Fast Fourier Transforms (FFTs) on the CM-5, we
calculated~Eq.~(\protect\ref{eq:2-pnt-correlation-fn}) via the
Wiener-Khintchin theorem \cite{Reif65}, first obtaining
the time-averaged magnitude squared of the Fourier
coefficients, from which
Eq.~(\protect\ref{eq:2-pnt-correlation-fn}) was obtained by an
inverse FFT.  In most cases, there was a substantial
region of exponential decay from which the two-point
correlation length~$\xi_2$ was obtained by a
least-squares fit of the form~$a \exp(-x/\xi_2)$; a
representative plot is given in
Fig.~\protect\ref{fig:2-point-corr-fn}.  The correlation
functions and corresponding values of~$\xi_2$ do not
change if Eq.~(\protect\ref{eq:2-pnt-correlation-fn}) is defined in
terms of the sign of the variables, ${\rm
sign}({y^t_i})$.

The two-point correlation functions decay more rapidly
as shown in Fig.~\protect\ref{fig:2-point-corr-mag-fn} if the
signs of the field values $y^{t}_i$ in
Eq.~(\protect\ref{eq:2-pnt-correlation-fn}) are replaced by their
magnitudes $|y^{t}_i|$.  The correlation
length~$\xi_2^{\rm mag}$ obtained from the initial
rapid exponential decay is approximately one lattice
spacing and changes little when the coupling constant
$g$ is varied over a large range, including near the
bifurcation point $g=g_c$.  For dimensionality~$d=1$,
we show below that this short length scale $\xi^{\rm
mag}_2$ is related to the dimension correlation length
$\xi_\delta$ and that both vary linearly with the
radius~$r$ of neighboring lattice sites that are
coupled together spatially.

As a possible alternative for characterizing the
spatial disorder of a nonlinear system, we also
calculated a correlation length~$\xi_I$ based on the
exponential decay of the mutual information
function~$I(\triangle{\bf X})$ \cite{mutual-info-refs}
of the variables~$y_i^t$.
Fig.~\protect\ref{fig:mutual-info-decay} shows the exponential
decay of a mutual information function~$I(\triangle{\bf
X})$ for two-dimensional square lattice of size~$L=256$
and for the parameter~$g=0.204$.  Again if the
magnitudes of the field variables are used when
calculating $I(\triangle{\bf X})$, the exponential
decay is much more rapid, with $\xi^{\rm mag}_I$ being
approximately one lattice spacing for a wide range in
$g$.

Although there is not yet a compelling theoretical
reason to prefer~$\xi_I$ over other correlation lengths
such as~$\xi_2$ \cite{Cr93}, the former has the
distinction of depending nonlinearly on the dynamics
and so may depend on details that are missed
by~$\xi_2$. For this reason, an increasing number of
scientists have reported correlation lengths in terms
of~$\xi_I$ \cite{Vastano88,vandeWater93,Bosch94}. As
shown in Fig.~\protect\ref{fig:mi-vs-2-point} for the
two-dimensional Miller-Huse CML on a square lattice,
the length scales~$\xi_2$ and~$\xi_I$ are linearly
related over a substantial dynamical range near the
ferromagnetic transition.  At least for the present
models, these lengths are equivalent measures of
spatial disorder and we report values only for~$\xi_2$
below.

\section{Results and Discussion}
\label{results}

In this section, we discuss our calculations of the
Lyapunov spectrum and of correlation lengths. Our goal
is to explore how spatial disorder (as characterized by
the two-point correlation length or by the mutual
information correlation length) is related to dynamical
complexity (as measured by the intensive dimension
density and by the dimension correlation length
Eq.~(\protect\ref{xi-delta-defn})) and to investigate how these
order parameters vary near the nonequilibrium
transition point~$g=g_c$ at which the magnetization
bifurcates to nonzero values
(Fig.~\protect\ref{fig:M-for-square-lattice}).

Results for the 2d~square lattice are given first,
followed by results for lattices with different
symmetries and dimensionalities.  We do not address
issues related to critical exponents of these different
models which have been discussed by Miller and Huse
\cite{Miller93} and more recently by Marcq and Chat\'e
\cite{Marcq95ising}.  Related interesting results on
similar CMLs have also recently been reported by
Boldrighini et al \cite{Boldrighini95}.

\subsection{Results for Two-Dimensional Square Lattices}
\label{2d-square-lattices}

For the two-dimensional periodic square lattice with
map Eq.~(\protect\ref{eq:Miller-Huse-map}),
Fig.~\protect\ref{fig:M-for-square-lattice} shows that there is a
bifurcation at~$g_c \simeq 0.205$. This bifurcation
corresponds to the onset of long-range order of the
lattice variables~$y_i^t$ as demonstrated by the
divergence of the two-point correlation length~$\xi_2$
as~$g$ approaches~$g_c$
(Fig.~\protect\ref{fig:divergence-of-xi_2}(a)). Over this same
parameter range, the dimension correlation
length~$\xi_\delta$ varies smoothly
(Fig.~\protect\ref{fig:divergence-of-xi_2}(b)), deviating by less
than four percent from a value of one lattice spacing
and attaining a maximum value close to where the
correlation length diverges. The Lyapunov
spectrum of exponents also varies smoothly from one
side of the transition to the other as shown in
Fig.~\protect\ref{fig:Lyapunov-spectra}. We conclude that chaotic
fluctuations have a short range, are decoupled from the
onset of long-range order measured by~$\xi_2$, and that
the spectrum of exponents is at most weakly dependent on the
onset of long-range spatial order.

To understand further how various dynamical quantities
change near the transition point, we have plotted in
Fig.~\protect\ref{fig:D-H-l1-vs-g-2d-square} the variation of
the Lyapunov fractal dimension density~$\delta$, of the metric
entropy density~$h$, and of the largest Lyapunov
exponent~$\lambda_1$ across the ferromagnetic
transition for a lattice of size~$L=16$, which is
already extensively chaotic according to
Fig.~\protect\ref{fig:ext-chaos}. As was the case for the
length~$\xi_\delta$ in
Fig.~\protect\ref{fig:divergence-of-xi_2}(b), these quantities
change by only a small amount through the transition
(at most by 20\%) and all go through a minimum close
to, but distinct from, the ferromagnetic transition
at~$g_c \approx 0.205$.
This result was surprising to us since one
consequence of coupling neighboring maps more strongly
(increasing the parameter~$g$) would intuitively be to
increase correlations between their dynamics, which
should decrease both~$\delta$ and~$h$.
Fig.~\protect\ref{fig:D-H-l1-vs-g-2d-square}(a) indicates that
roughly one quarter of the maximum number of degrees of
freedom
disappear when the lattice attains its minimum
dimension density of~$\delta \approx 0.746$.
(An upper bound of $\delta = 1$ is set by the integer
lattice spacing.)

It is not clear why the fractal dimension density and other
dynamical quantities have extrema near $g = 0.20$. For
an equilibrium non-dissipative CML of Ising dynamics on
a two-dimensional square lattice, Sakaguchi
\cite{Sakaguchi91Ising} did not
find a local minimum in the entropy~$H$ but instead
found a monotonic decrease consistent with the
analytical solution of the spin-$1/2$ Ising model on a
square lattice \cite{Landau80}. One explanation for the
extrema may be that the dissipative linear coupling in
Eq.~(\protect\ref{eq:the-cml}) becomes antidiffusive for~$g \ge
1/5$, enhancing rather than damping short-wavelength
fluctuations and so decorrelating nearby lattice
variables.

The issue is somewhat more subtle than this because the
existence of the minimum depends also on details of the
local map~$\phi(y)$ in Eq.~(\protect\ref{eq:the-cml}). For the less
chaotic lattice map Eq.~(\protect\ref{eq:slope-1.1-map}) with slope
of constant magnitude~1.1, an Ising-like transition
still occurs as shown by the bifurcation of the
magnetization near~$g_c \approx 0.168$ in
Fig.~\protect\ref{fig:M-for-square-lattice-map2}.
Fig.~\protect\ref{fig:D-H-l1-for-square-lattice-map2} now shows
that the Lyapunov dimension density~$\delta$ and
entropy density~$h$
decrease monotonically as the parameter $g$~is
increased, with the largest exponent~$\lambda_1$
remaining constant.

\subsection{Results for Other Lattices}
\label{other-lattices}

Figures~\ref{fig:M-for-square-lattice},
\ref{fig:divergence-of-xi_2}(a),
and~\ref{fig:D-H-l1-vs-g-2d-square} suggest that the
extrema of dynamical quantities may be related to the
ferromagnetic transition. On the other hand, the short
dimension correlation length in
Fig.~\protect\ref{fig:divergence-of-xi_2}(b) contradicts this by
implying that chaotic fluctuations occur over a length
scale that is short compared to the ferromagnetic
ordering. To understand this further, we have explored
CMLs of different symmetry and dimensionality and found
that the near-proximity of the extrema with the
transition is a coincidence for the two-dimensional
lattice with square symmetry.  More generally, the
positions of extrema seem to be determined simply by
the number of nearest neighbors~$n$, and not by the
symmetry or dimensionality of the CML or by the
position of the magnetization bifurcation point.

Fig.~\protect\ref{fig:M-for-hexagonal-lattice-map1} summarizes
calculations for a two-dimensional periodic hexagonal
lattice by plotting the dependence of magnetization~$M$
and of dimension correlation length~$\xi_\delta$ on the
coupling constant~$g$.  The magnetization bifurcates to
a nonzero value at~$g_c \approx 0.120$ which is a
smaller value than that for the square lattice since
the larger number of nearest neighbors (six versus
four) increases the effective strength of the diffusive
coupling.  The relative difference between the
transition at~$g=g_c$ and the positions of the extrema
in~$\xi_\delta$ and in related dynamical quantities is
substantially larger than was the case for the 2d
square lattice.  For the hexagonal lattice, extrema in
quantities like the length~$\xi_\delta$ occur at a
value close to~$g=1/(n+1) \approx 0.143$ where~$n=6$ is
the number of nearest neighbors.

A similar result is found for the same CML on a
3d~cubic lattice, as shown in
Fig.~\protect\ref{fig:M-for-cubic-lattice-map1}. The transition
at~$g \approx 0.11$ occurs at a value close to but
smaller than the value on the hexagonal lattice.
Extrema in the dynamical quantities like~$\xi_\delta$
again occur at a value close to~$1/(n+1)$ with~$n=6$.

That the positions of the extrema of dynamical
quantities is dependent primarily on the number of
nearest neighbors is a consequence of the
nearest-neighbor diffusive coupling in
Eq.~(\protect\ref{eq:the-cml}) and of the fact that, for the local
map Eq.~(\protect\ref{eq:Miller-Huse-map}), the chaos is
sufficiently strong to make the dimension correlation
length~$\xi_\delta$ quite small, about one lattice
length. That the positions of the extrema seem to be
given quantitatively by the specific
formula~$g=1/(n+1)$ is more delicate to understand but
may be related to the fact that the discrete Laplacian
operator changes from diffusive to antidiffusive
behavior at this value.  The value~$g=1/(n+1)$ is the
value for which the weight of each of the neighbors is equal
to the weight of the central lattice site to be updated.

For all CMLs that we studied, the dimension correlation
length~$\xi_\delta$ was about one lattice spacing in
size and this was also the ``radius'' of the diffusive
coupling in Eq.~(\protect\ref{eq:the-cml}). This suggests that the
length~$\xi_\delta$ may be determined by the spatial
extent of the diffusive coupling, becoming larger as
more sites are coupled to a given site.  This
conjecture was tested in one-space dimension by
coupling together, with equal weight~$g$, all lattice
variables within a radius~$r$ of a given site~$i$:
\begin{equation}
  y_i^{t+1} = \phi( y_i^t )
    + g \sum_{j=i-r}^{i+r} \left(
          \phi( y_j^t )
        - \phi( y_i^t )
      \right)
  .  \label{eq:radius-r-cml}
\end{equation}
For this one-dimensional periodic CML with the lattice
map Eq.~(\protect\ref{eq:Miller-Huse-map}), the dynamics varies in
a complicated way with radius~$r$.  For most initial
conditions, chaos was found for smaller radii ($r < 6$).
For larger radii~$r \ge 6$, the transients lasted much
longer and the asymptotic dynamics was periodic. As an
example, for~$r=10$ the dimension as a function of time
initially reached a value~$D=50$ even after $5000$
transient iterations were skipped; however, the dimension
then decreased
steadily to zero over the next 30,000 iterations. We
believe that this asymptotic periodic behavior is a
finite-size effect. For a sufficiently large system
size~$L$, with the crossover length increasing with the
radius~$r$, the asymptotic state should be chaotic.

For dimensionality~$d=1$, the dimension correlation
length~$\xi_\delta$ varies more strongly with
increasing radius~$r$ than with coupling constant~$g$,
which allows several different length scales to be
compared.  Fig.~\protect\ref{fig:1d-cml-results} shows that the
two-point correlation length $\xi_2^{\rm mag}$ obtained
using the magnitudes $|y_i^t|$ of the field values has
approximately the same linear dependence on the
coupling radius~$r$ as the dimension correlation
length~$\xi_\delta$.  In addition, these two length
scales are the same order of magnitude.  The two-point
correlation length $\xi_2$ obtained using the actual
field values is larger than $\xi_\delta$ and does not
have the same simple linear dependence on $r$. The
situation is then similar to results found for
spatiotemporal chaotic solutions of the 1d~periodic
CGLE \cite{Egolf95prl} in that the spatial correlation
length of fluctuations in the magnitude of a field
provides a way of estimating the length~$\xi_\delta$.

\section{Conclusions}
\label{conclusions}

In this paper, we have extended recent calculations
\cite{Egolf94nature,Egolf95prl,Egolf94thesis}
concerning the relation between spatial disorder and
dynamical complexity of a sustained homogeneous
nonequilibrium system from dimensionality~$d=1$ to
dimensionalities~$d=2$ and~$d=3$. This was accomplished
by choosing a coupled map lattice, Eq.~(\protect\ref{eq:the-cml}),
that underwent an Ising-like transition with diverging
two-point correlation length as a parameter~$g$ was
varied \cite{Miller93}. By comparing various length
scales such as the two-point correlation
length~$\xi_2$, the dimension correlation
length~$\xi_\delta$, and the two-point correlation
length of magnitude fluctuations~$\xi^{\rm mag}_2$ near
the transition point, we were able to show that the
lengths~$\xi_\delta$ and~$\xi^{\rm mag}_2$ were short,
of order one lattice spacing, even as the
length~$\xi_2$ diverged to infinity. In agreement with
previous work \cite{Egolf95prl}, the chaotic
fluctuations are decoupled from the average long-range
spatial order. The correlation length of magnitude
fluctuations seems to provide an effective way to
estimate the size of the length~$\xi_\delta$.

Our calculations of the Lyapunov spectrum and related
quantities such as the Lyapunov dimension density~$\delta$, metric
entropy density~$h$, and the largest Lyapunov
exponent~$\lambda_1$ show that the onset of long-range
spatial order (diverging~$\xi_2$) does not affect
dynamical invariants, which vary smoothly and weakly
through the transition point~$g=g_c$. Thus the average
spatial disorder (measured by~$\xi_2$) does not
determine dynamical complexity (measured
by~$\xi_\delta$). Rather surprisingly, the intensive
densities~$\delta$ and $h$
go through a minimum near the transition point so that
the onset of long-range order does not correspond to a
decrease in complexity. By examining CMLs of different
symmetry and of different dimensionality~$d$, we showed
that the positions~$g$ of the extrema were determined
by the number~$n$ of neighbors nearest to a given
lattice site (with $g \approx 1/(n+1)$) but not by the
symmetry or by~$d$.  This result can be understood as a
consequence of the extremely short dimension
correlation length~$\xi_\delta$, about one lattice
size, so that lattice variables are independent except
when they are nearest neighbors. We believe that the
minima in~$\delta$ and in~$h$ occur approximately when the
discrete Laplacian in Eq.~(\protect\ref{eq:the-cml}) becomes
antidiffusive with increasing parameter~$g$.
Short-wavelength fluctuations are then magnified
instead of damped, decreasing correlations between
neighboring sites.

Some of our results concerning extrema in dynamical
quantities have been independently obtained by
Boldrighini et al \cite{Boldrighini95} although these
authors worked with extensive, rather than intensive,
quantities and they did not determine whether their
calculations corresponded to extensively chaotic
regimes.
Boldrighini et al investigated CMLs of the form
Eq.~(\protect\ref{eq:the-cml}) for dimensionalities~$d=1$ and~$d=2$
but with some new local maps~$\phi(y)$. Besides also
finding extrema in the Lyapunov fractal dimension,
Boldrighini et al showed that the odd symmetry of the
map~$\phi(y)$ was not a sufficient condition for the
magnetization to bifurcate to a nonzero value.  Using a
strongly chaotic local map with slope of constant
magnitude equal to~5, they also showed that the
magnetization~$M$ did not bifurcate to a nonzero value
if the local map were made sufficiently chaotic
compared to the ordering caused by diffusion. In
Section~\protect\ref{other-lattices}, we used the less-chaotic local
map Eq.~(\protect\ref{eq:slope-1.1-map}) with slope of constant
magnitude~1.1 to show that the dimension density~$\delta$ can
decrease monotonically without a minimum even when the
magnetization~$M$ bifurcates to a nonzero value.  The
dependence of these minima on details of the local map
is not yet understood and should be pursued with
further studies.

The small values of $\xi_\delta$ in the present CMLs,
in the 1d~CGLE \cite{Egolf95prl}, and in a
nonequilibrium CML with algebraic decay of spatial
correlations \cite{OHern95algdecay} have several
interesting implications. One is that many previous
laboratory experiments \cite{space-time-expts} and
numerical simulations \cite{simulation-refs} concerning
spatiotemporal chaos are likely already extensive so
that it is meaningful to talk about dimension and
entropy densities (see also the earlier paper by Bohr
\cite{Bohr89}.) A second implication is that the
dimension correlation length~$\xi_\delta$ may not be a
useful order parameter for future studies of
spatiotemporal chaos since it depends only weakly on
parameters. A third implication is that the short value
of~$\xi_\delta$ suggests the nonexistence of
macroscopic chaotic states for dynamics with local
interactions, a point already made by several
researchers \cite{macroscopic-chaotic-states}. Finally,
we speculate that~$\xi_\delta$ is the length scale below
which one can replace chaotic fluctuations by a white
noise source when trying to develop a hydrodynamic
(long wavelength) description of spatiotemporal chaos
(see the discussion on pages 953-954 in
Ref.~\cite{Cr93}).

The short values of~$\xi_\delta$ raise the question of
what determines this length scale.  Our calculations on
the 1d~CML Eq.~(\protect\ref{eq:radius-r-cml}) with a variable
radius of coupling~$r$ suggest that the
length~$\xi_\delta$ is determined partly by the
length~$\xi^{\rm mag}_2$ characterizing magnitude
fluctuations although the reason for and the generality
of this correspondence is not understood
\cite{Egolf95prl}.  The length~$\xi_\delta$ is also
related to the radius~$r$ over which nearby lattice
sites are coupled together
(Fig.~\protect\ref{fig:1d-cml-results}). Further calculations
with different kinds local maps and of diffusive
operators and for different values of~$r$ should
provide further insight.

It is appropriate to finish with a discussion about the
relevance of these results for laboratory experiments.
As discussed at the end of Section~\protect\ref{exponents}, it does
not seem possible in the near future to calculate the
Lyapunov spectrum, the fractal dimension, or the
fractal dimension density of a high-dimensional
extensively chaotic experimental system for which only
time series measurements are available
\cite{time-series-limits}. Our success in calculating
these quantities was a result of having explicit
knowledge of the dynamical equations which could then
be integrated numerically on a powerful parallel
computer using algorithms whose complexity only grew
algebraically with the dimension~$D$
\cite{ref:alg-scaling}. For many laboratory
experiments, a quantitative mathematical description is
either lacking (e.g., for chemical reactions) or, if
known, is too difficult to work with numerically (e.g.,
the five three-dimensional Boussinesq equations
describing buoyancy-induced convection in a
large-aspect-ratio container \cite{Cr93}).

Our calculations in Section~\protect\ref{results} suggest that one
possible way to estimate the dimension correlation
length~$\xi_\delta$ may be to calculate the correlation
length of some function of the physical fields,
e.g., the field magnitude.  Another possibility will be
to discover and to validate algorithms that can
calculate the intensive dimension density
Eq.~(\protect\ref{delta-defn}) directly from time-series
measurements that are localized in space
\cite{density-from-time-series}, in lieu of calculating
a large extensive fractal dimension~$D$ and then
dividing by the extensive system volume~$V$.  Several
steps have been taken in this direction
\cite{density-from-time-series}, but a theoretical
foundation has not yet been established nor have the
numerical algorithms been adequately tested.

\section*{Acknowledgements}

We thank L.~Bunimovich, H.~Chat\'e, P.~Hohenberg,
D.~Huse, H.~Riecke, and J.~Socolar for useful
discussions. This work was supported by grants
NSF-CDA-91-23483 and NSF-DMS-93-07893 of the National
Science Foundation, by grant DOE-DE-FG05-94ER25214 of
the Department of Energy and by an allotment of CM-5
computer time at the National Center for Supercomputing
Applications.  The second author (D.A.E.) would like to
thank the Office of Naval Research for fellowship
support. Much of the research reported here was carried
out by the first author as part of his undergraduate
honors thesis in physics at Duke University.

\newpage

\begin{figure}
\caption{
Lyapunov fractal dimension~$D$, Eq.~(\protect\ref{eq:dim-defn}),
versus iteration number~$t$ for a periodic 2d~CML on a
square lattice with local map
Eq.~(\protect\ref{eq:Miller-Huse-map}), with coupling
constant~$g=0.202$ and with lattice size $L=16$.
Fluctuations in~$D$ damp out slowly with increasing
time, giving a relative accuracy of about 0.1\% in the
dimension (here estimated to be $D = 191.2$).  }
\label{fig:dim-vs-time}
\end{figure}

\begin{figure}
\caption{ Lyapunov fractal dimension~$D$ versus the
number of lattice points~$N=L^2$ (which is also the
system volume) for a 2d periodic CML on a square
lattice with local map Eq.~(\protect\ref{eq:Miller-Huse-map}) and
coupling constant~$g=0.199$.  Each dimension value was
obtained from a plot similar to
Fig.~\protect\ref{fig:dim-vs-time} over a time scale
of~$T=50,000$ iterations. {\bf (a)} The Lyapunov
dimension increases linearly and so extensively
with~$N$ for~$N > 100$. A least-squares fitted line
through the four right-most points gives a Lyapunov
dimension density (slope) of~$\delta = 0.746$ and an
intercept of~$0.0077$.  {\bf (b)} The normalized
deviation $(D/L^2 - \delta)/\delta$ of the
dimension density~$\delta$ from
the dimension density predicted from the dimension~$D$
per volume~$L^2$ illustrates the rapid onset of
extensive chaos for small system sizes~$L > 9$. }
\label{fig:ext-chaos}
\end{figure}

\begin{figure}
\caption{
Average magnetization~$M$
(Eq.~(\protect\ref{eq:magnetization-defn})) versus the lattice
coupling strength~$g$ for a two-dimensional CML with
local map Eq.~(\protect\ref{eq:Miller-Huse-map}) on a square
lattice of size~$L=128$ after an integration time
of~$T=3 \times 10^5$.  The magnetization bifurcates
from a nonzero value at the critical value $g_c
=0.205$. }
\label{fig:M-for-square-lattice}
\end{figure}

\begin{figure}
\caption{
Log-linear plot of the exponentially decaying two-point
correlation function Eq.~(\protect\ref{eq:2-pnt-correlation-fn}) of
the two-dimensional CML Eq.~(\protect\ref{eq:the-cml}) with local
map
Eq.~(\protect\ref{eq:Miller-Huse-map}) on a square lattice
for lattice sizes of~$L=64, \cdots, 1024$.  The
coupling constant was~$g=0.199$ and the integration
time was~$T=3 \times 10^5$. A reasonably accurate
estimate of the correlation length~$\xi_2 \simeq 6$ is
found only for sizes~$L \ge 256$.  The arrows indicate
the range over which a least-squares linear fit was
used to extract the length~$\xi_2$. }
\label{fig:2-point-corr-fn}
\end{figure}

\begin{figure}
\caption{
Plot of the two-point correlation function
Eq.~(\protect\ref{eq:2-pnt-correlation-fn}) for the magnitudes
$|y^t_i|$ of the field for the 2d periodic CML with
local map Eq.~(\protect\ref{eq:Miller-Huse-map}) on a square
lattice with~$g=0.200$.  A system size $L=64$ and an
integration time of $T=5 \times 10^5$ were used.  The
correlation length $\xi_2^{\rm mag} \approx 1.4$ is
smaller than the length~$\xi_2 \approx 6$ based on the
values~$y^t_i$. }
\label{fig:2-point-corr-mag-fn}
\end{figure}

\begin{figure}
\caption{
Log-linear plot showing the exponential decay of the
mutual information function~$I(\triangle{x})$ as a
function of the spatial separation~$\triangle{x}$ for
the 2d~periodic CML on a square lattice with local map
Eq.~(\protect\ref{eq:Miller-Huse-map}), coupling
constant~$g=0.204$, system size~$L=256$, and effective
integration time~$T=128,000$ (following a transient
time of~$100,000$). The arrows indicate the range over
which a linear least-squares fit was used to calculate
the correlation length~$\xi_I \approx 11.7$.  A simple
binning procedure with 32~bins was adequate so that the
more sophisticated algorithms of Fraser et al
\protect\cite{mutual-info-refs} were not needed. }
\label{fig:mutual-info-decay}
\end{figure}

\begin{figure}
\caption{
Two-point correlation length~$\xi_2$ versus the mutual
information correlation length~$\xi_I$ for a 2d
periodic CML on a square lattice with local map
Eq.~(\protect\ref{eq:Miller-Huse-map}). Lattices sizes of~$L=1024$
and~$L=256$ and integration times of~$T=300,000$
and~$T=128,000$ were used respectively when
calculating~$\xi_2$ and~$\xi_I$. The points correspond
to the parameter range $0.1900 \le g \le 0.2045$ with the
smaller~$g$ values occurring to the left. The values
of~$\xi_2$ and~$\xi_I$ were obtained from plots similar
to Fig.~\protect\ref{fig:2-point-corr-fn} and
Fig.~\protect\ref{fig:mutual-info-decay} }
\label{fig:mi-vs-2-point}
\end{figure}

\begin{figure}
\caption{
{\bf (a)} Divergence of~$\xi_2$ for the 2d periodic CML
on a square lattice with local map
Eq.~(\protect\ref{eq:Miller-Huse-map}).  The lattice size
was~$L=1024$ and the integration time was~$T=3 \times
10^5$ based on an ensemble average over 64
realizations. The correlation length diverges with
critical exponent~$\approx -1.0$ at the extrapolated
point~$g=0.2054$ which coincides within numerical
accuracy with the pitchfork bifurcation of the
magnetization.
{\bf (b)} The dimension correlation length~$\xi_\delta$
(Eq.~(\protect\ref{xi-delta-defn})) versus parameter~$g$ for the
same parameter values, based on dimension densities
calculated from extensive chaos curves like
Fig.~\protect\ref{fig:ext-chaos}. This length scale is roughly
one lattice size, varies smoothly with~$g$, and attains
a maximum value for~$g \approx 0.200$. }
\label{fig:divergence-of-xi_2}
\end{figure}

\begin{figure}
\caption{
Lyapunov exponents~$\lambda_i$ versus the intensive
index~$x=i/L^2$ for the 2d periodic CML on a square
lattice with local map Eq.~(\protect\ref{eq:Miller-Huse-map}) for
the lattice size~$L=24$, for an integration
time~$T=50,000$, and for lattice coupling
values~$g=0.19$ and~$g=0.22$ which straddle the
transition at~$g=0.205$. The exponents form an
approximately continuous function that changes smoothly
with parameter~$g$. Only the first $512$ of the~$576$
possible exponents are shown. }
\label{fig:Lyapunov-spectra}
\end{figure}

\begin{figure}
\caption{
{\bf (a)} The Lyapunov fractal dimension density~$\delta$,
{\bf (b)} the entropy
density~$h$, and {\bf (c)}
the largest Lyapunov exponent~$\lambda_1$ versus the
coupling constant~$g$ in the vicinity of the
ferromagnetic transition of~$g_c=0.205$ according to
the magnetization curve in
Fig.~\protect\ref{fig:M-for-square-lattice} for the 2d CML on a
square lattice with local map
Eq.~(\protect\ref{eq:Miller-Huse-map}). System sizes of at
least~$L=16$ were used
with a fixed integration time
{}~$T=5 \times 10^4$ and an ensemble average
over~$p=64$ systems. The quantities~$\delta$ and~$h$
have local minima (although at slightly different~$g$
values) to the left of where the magnetization
bifurcates. The error bars of each data point are about
the size of the points themselves except for those
of~$\lambda_1$ which are much larger.  }
\label{fig:D-H-l1-vs-g-2d-square}
\end{figure}

\begin{figure}
\caption{
Magnetization $M$ (Eq.~(\protect\ref{eq:magnetization-defn}))
versus coupling strength~$g$ for the 2d periodic CML on
a square lattice but with the less chaotic lattice map
given by Eq.~(\protect\ref{eq:slope-1.1-map}). Other parameter
values were the sizes, $L=64$ and~$L=128$ and the total
integration time, $T=3 \times 10^5$.  The bifurcation
occurs at approximately~$g_c \approx 0.168$.  }
\label{fig:M-for-square-lattice-map2}
\end{figure}

\begin{figure}
\caption{
Lyapunov dimension density~$\delta$ and entropy
density~$h$ versus coupling constant~$g$ for the
2d~periodic CML with local map
Eq.~(\protect\ref{eq:slope-1.1-map}), for lattice size $L=32$ and
for an integration time $T=10^4$. The largest Lyapunov
exponent~$\lambda_1 = 0.095$ remains constant in the
range $0.160 < g < 0.180$ and so is not plotted. For
this substantially less chaotic local map, extrema in
the dynamical quantities do not occur.  }
\label{fig:D-H-l1-for-square-lattice-map2}
\end{figure}

\begin{figure}
\caption{
{\bf (a)} Magnetization~$M$
(Eq.~(\protect\ref{eq:magnetization-defn})) versus coupling
strength~$g$ for 2d periodic CML on a hexagonal lattice
with local map Eq.~(\protect\ref{eq:Miller-Huse-map}), with
integration time~$T=1 \times 10^5$ time units, for a
system size $L=64$, and for an ensemble average
of~$p=64$ elements.  A ferromagnetic transition occurs
at~$g=g_c \approx 0.12$.
{\bf (b)} Dimension correlation length~$\xi_\delta$
versus the lattice coupling constant~$g$ for the same
parameters except with a system size $L=16$ and
integration time $T=5 \times 10^4$.  The maximum
in~$\xi_\delta$ does not coincide with the bifurcation
of the magnetization in {\bf (a)}. Over this range, the
Lyapunov dimension decreases from a value of~106 to a
minimum of~$D \approx 100$ at~$g=0.14$.  }
\label{fig:M-for-hexagonal-lattice-map1}
\end{figure}

\begin{figure}
\caption{
{\bf (a)} Magnetization~$M$
(Eq.~(\protect\ref{eq:magnetization-defn})) versus lattice coupling
strength~$g$ for the 3d~periodic CML with local map
Eq.~(\protect\ref{eq:Miller-Huse-map}) on a cubic lattice. A
system size of $L=16$, a total integration time
of~$T=1 \times 10^5$ time units, and an ensemble
average of~$p=64$ realizations were used.  The
magnetization bifurcates to a nonzero value at~$g=g_c
\approx 0.11$.
{\bf (b)} The dimension correlation length~$\xi_\delta$
versus~$g$ for the same parameters, calculated on a
lattice of size~$L=6$ and for a time of $T=1 \times
10^4$ time units.  The corresponding dimension~$D$
varies from 120 down to a minimum of about~$105$
at~$g=0.15$.  }
\label{fig:M-for-cubic-lattice-map1}
\end{figure}

\begin{figure}
\caption{
Variation of the correlation length~$\xi_2$, the
correlation length of magnitude fluctuations $\xi^{\rm
mag}_2$, and the dimension correlation
length~$\xi_\delta$ with the radius of spatial
coupling~$r$ for the 1d~periodic CML
Eq.~(\protect\ref{eq:radius-r-cml}) with local map
Eq.~(\protect\ref{eq:Miller-Huse-map}). The system sizes ranged
from~$L=64$ to~$L=1024$ with larger systems being used
for larger radii. The integration time was~$T=300,000$
for all systems.  The length $\xi_\delta \approx 5.3$
for~$r=5$ is substantially larger than the maximum
value of~$\xi_\delta
\approx 1.2$ found for the 2d Miller-Huse CML with nearest
neighbor coupling.
}
\label{fig:1d-cml-results}
\end{figure}


\begin{references}
\bibitem{ref:alg-scaling}
For a numerical dynamical model with~$N$ independent
variables, an optimal integration scheme will advance
the variables in one time step with a computational
effort that grows at best as slowly as~$O(N)$. Each
linearized equation also involves~$N$ variables and so
the effort needed to calculate~$K$ Lyapunov exponents
by the algorithm discussed in Section~\protect\ref{methods} over a
fixed time time~$T$ will grow at best as slowly as~$O(T
K N)$.  Since the number of exponents needed to
estimate the Lyapunov dimension grows linearly with
system volume, as does the number~$N$ itself, we
conclude that the computational effort to calculate the
Lyapunov dimension for an extensively chaotic system,
for an observation time~$T$, will grow at best as
slowly as~$O(T N^2)$, i.e., algebraically with~$N$. It
is presently not understood how the total integration
time~$T$ should itself scale with system volume or
with~$N$ and so the exponent of the algebraic scaling
with dimension~$D$ is not yet known \cite{Egolf95ics}.
\end{references}
\end{document}